\documentclass[]{aa}
\begin{document}

  \thesaurus{06     
              (12.05.1;  
   	       11.01.2;  
	       11.11.1)}	 
   \title{The nature of the extreme kinematics in the extended gas of high redshift radio
galaxies.}

\author{M.Villar-Mart\'\i n
          \inst{1}
          L. Binette 
	  \inst{2}
	  \and
	  R.A.E. Fosbury
	  \inst{3}
	}

   \offprints{M.Villar-Mart\'\i n. email: villard@iap.fr}

   \institute{Institute d'Astrophysique de Paris (IAP),
	98 bis Bd Arago, F75014 Paris, France 
	\and
	Instituto de Astronom\'\i a, UNAM, Apartado Postal 70-264, D.F. 04510,
	Mexico
    	\and 
	Space Telescope European Coordinating Facility, Karl Schwarschild Str. 2, D-85748, Garching, Germany
	}	
   \date{}

\authorrunning{Villar-Mart\'\i n {\it et al.}}
\titlerunning{Kinematics of the extended gas in HZRG.}

   \maketitle

\begin{abstract}

	We present UV rest frame spectra of 3 power\-ful narrow line radio galaxies  
and the hyperluminous type 2 active galaxy SMM02399-0136, all at  high redshift ($z>$2).
 We find high velocities  (${\rm FWHM}>1000$\, km s$^{-1}$) in the {\it extended} gas of {\it all} objects.
A natural explanation  is the interaction
between the radio jet and the ambient gas, that drives shocks into the
gas and accelerates the clouds. However,
the existence of high velocities in regions where such interactions are 
not taking place
implies that other processes can play a
role. We discuss here several possible mechanisms.

\end{abstract}

\section{Introduction}

The existence of high velocities (FWHM$>$1000 km s$^{-1}$) (McCarthy
{\it et al.} 1996) in the extended gas (EELR) of high redshift ($z>$2)
radio galaxies (HZRG) is in contrast with the more relaxed kinematics
observed in the majority of low redshift radio galaxies (FWHM$<$400 km
s$^{-1}$) (Tadhunter {\it et al.} 1989). The nature of such extreme
kinematic motions is not well understood. We investigate here this
issue studying the kinematics of the extended gas in a small sample of
4 distant active galaxies ($z>$2): MRC1558-003, MRC2025-218 and
MRC2104-242 (radio galaxies) and SMM02399-0136 (hyperluminous type 2
active galaxy with very weak radio emission).

\section{Observations and data reduction}

The spectroscopic observations were carried out on the nights 1997 July 3-5
and 1998 July 25-27
using  the EMMI multi-purpose instrument at the NTT (New Technology
Telescope) in La Silla Observatory
(ESO-Chile).
 The detector was a  Tektronix CCD with 2048$\times$2048 pixels of size 24 $\mu$m,
 resulting in a spatial scale of 0.27 arcsec per pixel.  We used EMMI in RILD
spectroscopic mode (Red Ima\-ging and Low Disperson Spectroscopy). We used 
the same grism  (\#3)  for
all objects. This has a blaze wavelength of 4600 \AA,
dispersion
5.9 \AA/pixel and wavelength range 4000-8300 \AA.
The  slit was aligned with the radio axis for the three radio galaxies.
We positioned the slit along the two main optical components of SMM02399-0136
 (L1 and L2, adopting
the nomenclature of
Ivinson {\it et al.} 1998 [IV98 hereafter]). 
A log of the spectroscopic observations is shown in Table 1.

 \begin{table*}
\normalsize
\centering
\caption{{\it Log of the spectroscopic observing run}. The redshift (calculated from the 
spatially integrated Ly$\alpha$ emission) is also indicated.}
\begin{tabular}{cccccccc}
\hline

 Object	& $z$ & Date &  $t_{exp}$  & Slit & Average &    Resol.(\AA) & PA  \\	
	&	&	& (sec)		&  (")	& Seeing(")	& at 
$\lambda_{Ly\alpha}^{obs}$   & of slit	\\ \hline
MRC2104-242	&  2.492	& 3/Jul/97	& 4$\times$1800	& 1.5	& 1.10	& 10.4	& 22	\\
MRC2025-218	& 2.632	& 3/Jul/97	& 3$\times$1800	& 1.5	& 1.15	& 10.9  & 30	\\
MRC1558-003	& 2.530	& 25-26/Jul/98	& 4$\times$1800	& 1.5	& 1.00	& 10.3  & 75	\\
SMM02399-0136	& 2.803	& 25-26/Jul/98	& 5$\times$1800	& 1.5	& 1.20  & 10.3	& 88.6	\\
\hline	
\end{tabular}
\end{table*}

Standard  data reduction techniques were applied using IRAF software (see Villar-Mart\'\i n 
{\it et al.} 1998 for a more detailed description). 

\section{The fitting procedure}

	In order to study the kinematics of the gas, we fitted the emission
lines with a Gaussian profile at every spatial position (pixel).
Several spatial pixels were added where the emission was too
faint. We used the Starlink packa\-ge DIPSO for this purpose.
The FWHM, flux and central
wavelength  were measured from the 
Gaussian fitted to the line profile.  The FWHM was
corrected in quadrature for instrumental broadening (the instrumental profiles in the observed frame
are given in Table 1).

Single Gaussians did not always provide a perfect fit and underlying
broad wings were sometimes present. This is probably due
to the presence of several kinematic components and/or absorption of
Ly$\alpha$ by neutral hydrogen. To elimi\-nate uncertainties due to the second
mechanism, we also present the result of the fit for the second strongest
emi\-ssion line CIV$\lambda$1550 not susceptible of hydrogen absorption.

At
the s/n of the data, we are confined to using single Gaussian fits to the
lines, a procedure which is the same as that followed by Tadhunter {\it et
al.} (1989).
 A single Gaussian fit will a) lose
any information about multiple components  b) neglect 
 any possible weak broad
under\-lying wings (as is observed in MRC2104-242, see Fig.~1 left cloud),
since the fit will be optimized for the dominant part of the line. 
However, the main goal of this paper does not require such a precise
analysis. The broad wing on MRC2104-242 will have little influence on the fit, which will be
dominated by the strong, narrower component.

\section{Results}

 \begin{figure} 
 \includegraphics{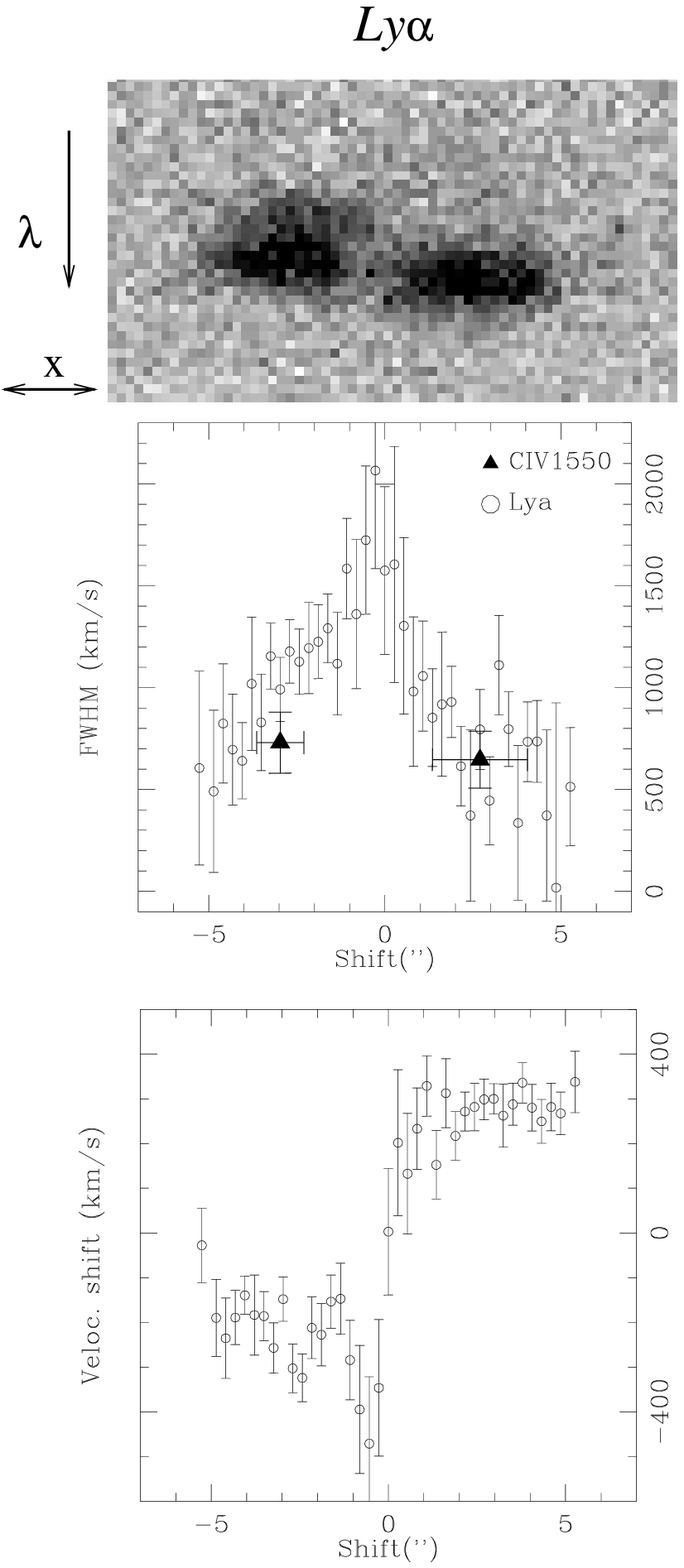}
\vspace{6.2in}
\caption{{\it MRC2104-242.}  Upper panel: 2-D spectrum of the
Ly$\alpha$ spectral region with the dispersion in $\lambda$ running
vertically.  Middle: Spatial variation of the FWHM. Lower panel:
Velocity shift with respect to the emission in the center of symmetry
of the Ly$\alpha$ distribution. The spatial shift has been calculated
with respect to the same position.  The horizontal lines indicate
the pixels added along the spatial direction. The three panels have
the same spatial scale and are aligned so that vertical lines connect
the same spatial positions in the slit. The errorbars (vertical lines)
include the
errors of the fit, as well as the error in the measurement of
the instrumental profile.}
\end{figure} 
	
 \begin{figure} 
\includegraphics{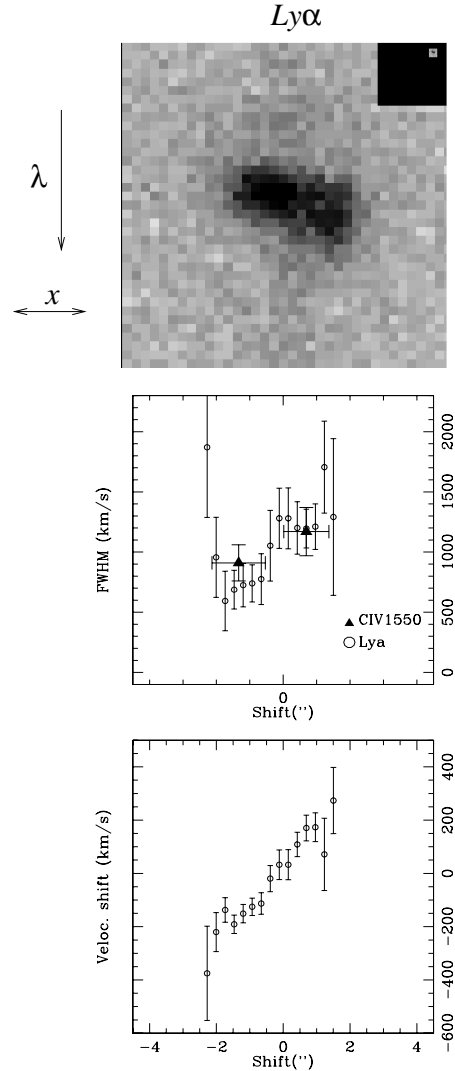}
\vspace{6.2in}
\caption{{\it MRC2025-218.} Panels and symbols as in Fig.~1. 
The spatial and velocity shifts have been calculated with respect
to the centroid of continuum emission. }
\end{figure}
 
 \begin{figure} 
\includegraphics{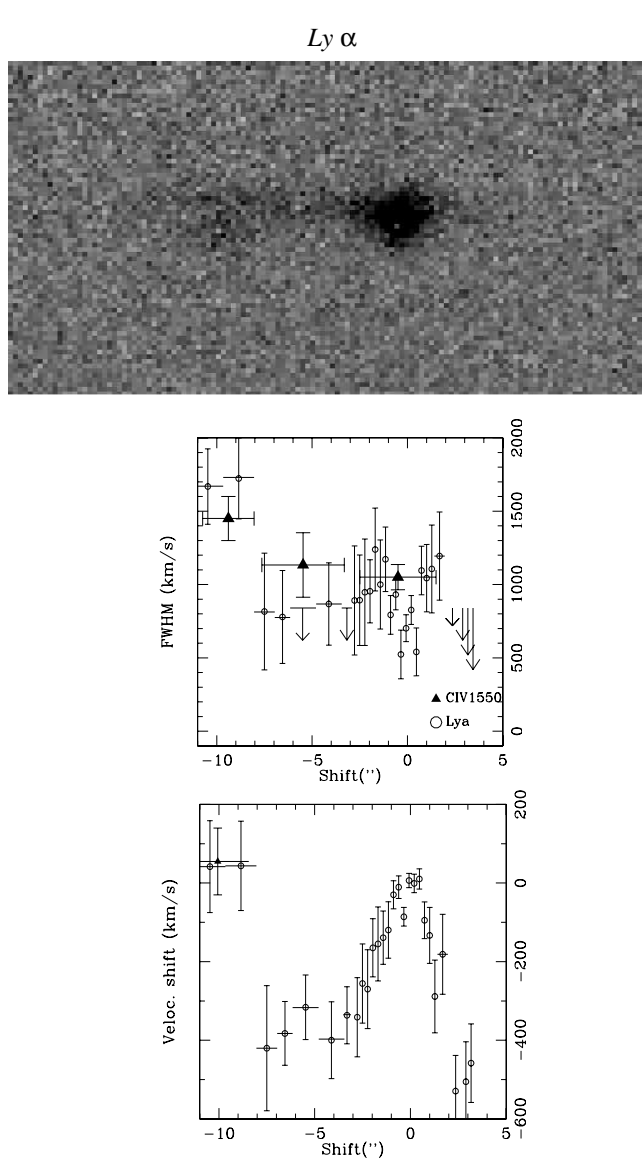}
\vspace{6.2in}
\caption{{\it MRC1558-003.} Panels and symbols as in Fig.~1. 
The spatial and velocity shifts have been calculated with respect
to the nuclear emission. 
The high velocity component 10 arc sec from the $x=$0 position is  located beyond the
radio structures.}
\end{figure} 	

 \begin{figure} 
\includegraphics{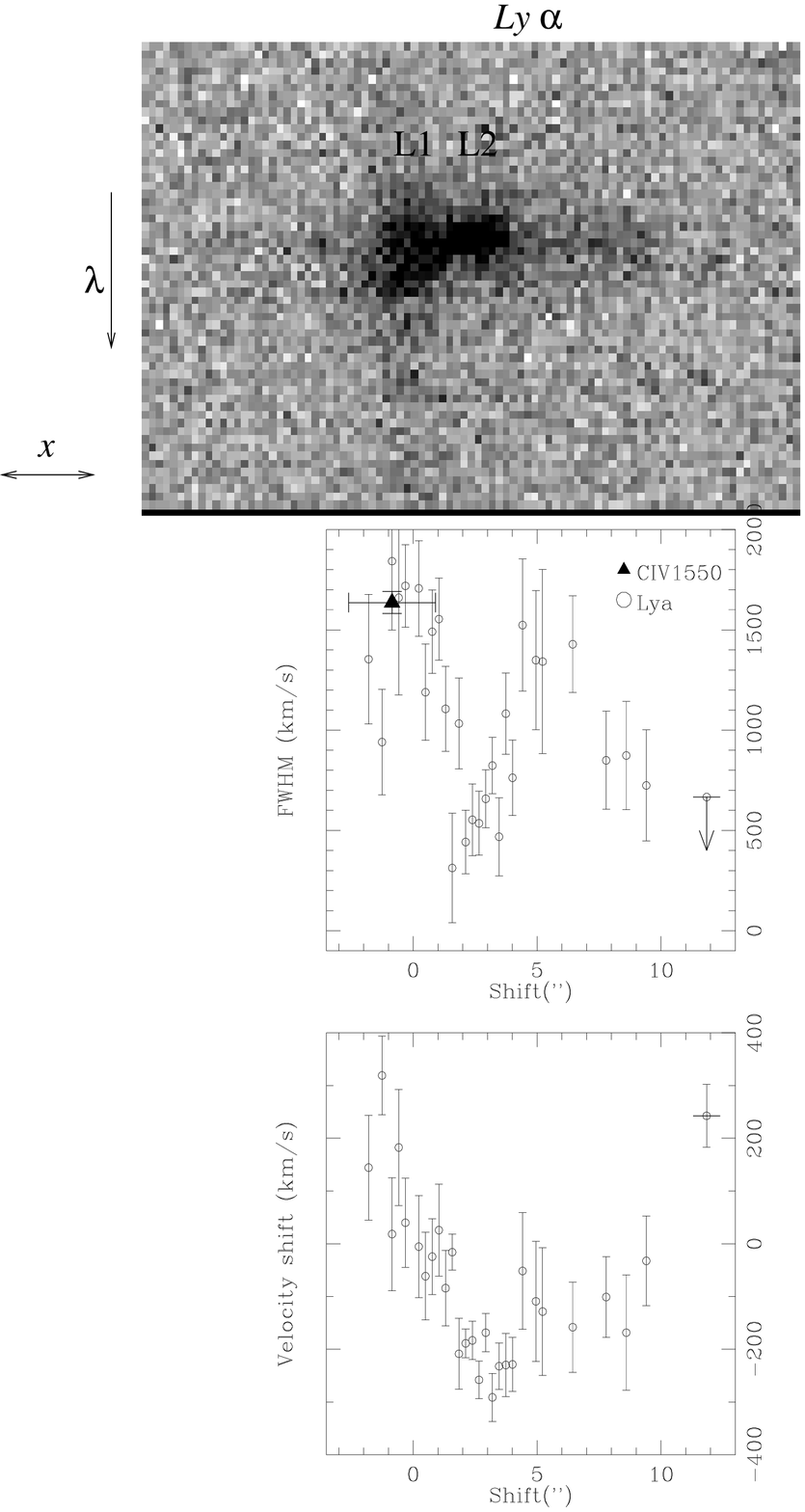}
\vspace{6.2in}
\caption{{\it SMM02399-0136.} Panels and symbols as in Fig.~1. The two 
components of the system (IV98) have been indicated (L1 and L2).
The spatial and velocity shifts have been calculated with respect
to the centroid of continuum emission. Note the presence of large velocities in the extended gas of
 this very weak radio  source. }
\end{figure}

	We present in Figs. 1 to 4 the results of our analysis for
the 4 targets in the sample. The upper panel in each figure is the 2-D
spectrum of the Ly$\alpha$ spectral region with the dispersion in $\lambda$
running vertically. The middle panel presents
the spatial variation of the FWHM and the bottom panel shows the spatial
variation of 
the velocity shift of the Ly$\alpha$ emission (open circles) across the nebula. 
The 3 panels in each figure have the same spatial scale and are aligned
so that vertical lines join the same spatial positions. 
FWHM for CIV$\lambda$1550 is also plotted
for comparison (solid triangles). 

\begin{itemize}

\item MRC2104-242 (Fig.~1): 
Ly$\alpha$ shows a bimodal distribution and is extended over 
$\sim$12 arcsec
along the slit aligned with  the radio axis. The two blobs lie in between the radio lobes
(McCarthy {\it et al.} 1990). The bimodal distribution is also apparent in 
CIV.  The two clumps present high FWHM values (1100 km s$^{-1}$ and 900 km s$^{-1}$ 
respectively, for the spatially integrated spectra) and are shifted by $\sim$500 km s$^{-1}$
(consistent with McCarthy {\it et al.} 1990, Koekemoer {\it et al.} 1996).
Kinematic substructure is observed in the two blobs.
 The velocity curve (Fig.~1, bottom panel)  is
rather flat across each blob. CIV is also extended and presents high FWHM values
($\sim$700 km s$^{-1}$, see Fig.~1).

\item MRC2025-218 (Fig.~2): Ly$\alpha$ shows a bimodal
distribution and is extended over $\sim$4.5 arc sec. This structure
lies between the radio lobes (Pentericci {\it et al.} 1998).
Continuum is detected in our spectra. Large and rather constant FWHM
values are measured across the two components ($\sim$1200 km s$^{-1}$
and 700 km s$^{-1}$ respectively).  The velocity curve is rather steep
across the nebula varying smoothly over a range of $\sim$600 km s$^{-1}$.  CIV is
also extended and presents similar FWHM as Ly$\alpha$.

\item MRC1558-003 (Fig.~3): Ly$\alpha$ is extended over $\geq$15 arc
sec. A bright component is detected as well as diffu\-se, very
extended emission which show large velocity widths at $\sim$10 arc sec
from the main component. The optical (R\"otggering {\it et al.} 1994)
and radio astrometry (R\"ottgering {\it et al.} 1996, Rhee {\it et
al.} 1996) locate the high velocity region several arcsec {\it beyond} the radio
structures. CIV is also extended and presents very large FWHM
within the high velocity region, consistent with the Ly$\alpha$
measurement. There is no apparent pattern in the velocity curve of the
ionized gas.

\item SMM02399-0136 (Fig.~4): This object is a hyperluminous active
galaxy, gravitationally lensed by a foreground cluster (IV98).  It consists
of two main optical sources L1 and L2. The radio emission is very
weak, below the detection tresholds of most radio surveys.  
The radio, submm and optical properties are consistent with a
scenario where L1 contains an active nucleus and L2 is an
interacting companion. The system is undergoing strong starburst activity.
 We detect Ly$\alpha$ emission across $\sim$20
arc sec. Two main components are revealed by our spectra, coincident
with (L1 and L2).  Ly$\alpha$ is relatively broad in L1 (FWHM$\sim$1800 km
$s^{-1}$) and narrower in L2 (FWHM$\sim$300-700 km s$^{-1}$). A high velocity
region (FWHM$\sim$1500 km $^{-1}$) is detected at the border of L2. A
spectrally unresolved region lies at $\sim$12 arc sec from L1.

\end{itemize}

	The four objects present complex kinematics and show
that there is a large variety of kinematic behaviour in high redshift active
 galaxies. All objects present certain common characteristics:

\begin{itemize}

\item High velocities (FWHM$>$1000 km s$^{-1}$) in the {\it extended gas}

\item Velocity shift of
the Ly$\alpha$ emission across the nebula varying over a range $<$700 km 
s$^{-1}$,  although the velocity curves are
rather different from object to object. Similar values are observed
in low redshift radio ga\-laxies (Tadhunter {\it et al.} 1989) 

\item  Presence of at least 
two different kinematic components which seem to be spatially distinct. Such components look like individual
clumps in the case of MRC2025-218, MRC2104-242 and SMM02399-0136.
Diffuse and fainter line emission is present in the spectra of SMM02399-0136
and MRC1558-003. Narrow band Ly$\alpha$ narrow band images   show also
diffuse Ly$\alpha$ emission in MRC2025-218 and MRC2104-242 in addition to
the brightest components (McCarthy 
{\it et al.}  1990, Pentericci {\it et al.} 1998).

\end{itemize}

\section{Discussion}

High velocities have been observed in the EELR of many HZRG (McCarthy
{\it et al.} 1996). The alignment between the radio and optical
structures (McCarthy {\it et al.} 1987, Chambers {\it et al.} 1987)
and the anticorrelation between the size of the radio source and the
velocity dispersion found for HZRG (van Ojik 1995)
suggest that the jet is interacting with the ambient gas. Studies of
radio gala\-xies at intermediate redshift with clear signs of such
interac\-tions as well as hydrodynamical simulations show that this
process can produce large FWHM ($>$1000 km s$^{-1}$) ({\it e.g.} Villar-Mart\'\i n {\it et al}
1999, Clark {\it et al.} 1997).  Some HZRG  show clear
evidence
for jet-cloud interactions  ({\it e.g.} van Ojik et
al. 1996) and this process is surely having an effect in
some high redshift radio galaxies. It could be also the case of MRC2025-214 and MRC2104-242,
which show the alignment effect (McCarthy {\it et al.} 1990) and large
line widths {\it inside} the radio structures. However, we have also
measured high velocities in the EELR of MRC1558-003 {\it beyond} the
radio structures (Fig.~3) and the extended gas of the galaxy-galaxy
interacting system SMM02399-236.  Bremer {\it et al.} (1992) reported
the detection of extended Ly$\alpha$ emission in the radio quiet
quasar 0055-264 ($z=$3.66) with FWHM$\sim$1000 km s$^{-1}$. Jet cloud
interactions cannot explain the extreme kinematics in these
objects. Another accelerating mechanism is at work which could also
play a role in many other HZRG.

IV98 have proposed that
SMM02399-0136 is a system in which two  companions are interacting: 
L1 (that contains an active
nucleus) and  L2. The interaction  has induced starburst  activity
responsible for the submm and weak radio emissions. Another possibility is that the
radio emission is due to a frustrated radio jet, since the steep radio spectrum is consistent
with an AGN origin (IV98). The radio emission is extended along
PA71, close to PA88.6 (L1-L2 direction) and the projected size ($\sim$7.9 arc sec)
is  larger than the L1-L2 extension ($\sim$4 arc sec). Therefore, the high velocity gas detected
beyond L2  is probably inside the radio structures. In this case, the interaction between the radio
jet and the ambient gas could be responsible for the large FWHM values measured beyond L2.

However, if SMM02399-0136 is a system of two interaction companions with strong starburst
activity
[as is the case for many $\mu$Jy radio sources (Lowenthal 1997)],
jet-cloud interactions cannot explain the gas
kinematics in SMM02399-0136.
\footnote{SMM02399-0136  is gravitationally lensed
by a foreground cluster. The appropriate geometry with respect to the
lensing source could produce a greater distortion of component L1
due to its compact morphology. L1 emission might extend beyond L2
with the result that the high velocities measured in L1 could be
contaminating measurements of the extended gas.  We would then
 expect the same effect for NV and CIV, which are strongly nucleated
and have similar fluxes in L1 than Ly$\alpha$ (see Fig.~3 in IV98).
However, these lines are detected only in L1.}

We discuss here several possible mechanisms that could explain the extreme kinematics in some
HZRG.

\begin{itemize}

\item{\it Broad scattered lines.}

Many distant radio galaxies ($z>$2) show polarized continuum  with the electric
vector perpendicular to the axis of the optical (UV rest frame)
structures  (Cimatti
{\it et al.}  1998, Fosbury {\it et al.} 1998b,1999). 
This is consistent with an scenario in which powerful radio gala\-xies contain
QSO nuclei whose FUV emission we see scattered by extended dust structures.
 The
emission from the broad line region should also  be scattered and
therefore broad
lines could be detected within the extended gas.  
  Scattering preserves the equivalent width of BLR lines against
the nuclear continuum and if this mechanism dominated, we should measure
similar EW in the extended gas.  We have measured a lower limit for the
EW of the CIV$\lambda$1550 line (not affected
by neutral hydrogen absorption). We obtain EW(CIV)$\geq$100, which is quite
large compared
with typical values measured in quasars at high redshift 
(Corbin \& Francis 1994). This suggests that the line is dominated by direct
light, rather than scattered.

	This mechanism is not likely to
play a role in radio galaxies in general. In effect, NIR spectroscopy of distant
narrow line radio galaxies ($z=$2.2-2.6) reveals FWHM$>$1000 km
s$^{-1}$ for both permitted and {\it forbidden} lines (Evans
1998).  Eventhough the spectra of HZRG are  customarily integrated
along the spatial dimension, 
the extended emission frequently domi\-nates the  integrated spectra
 ({\it e.g.}
McCarthy {\it et al.} 1996, Stockton {\it et al.} 1996) and therefore
the observed large velocities could well originate from the
extended regions themselves. At $z\sim$1 the [OII] emission has simi\-larly
 revealed high velocities
(FWHM$>$1000 km s$^{-1}$) within the {\it extended} gas regions
 of some radio
galaxies (McCarthy {\it et al.} 1996).

\item{\it Infall.}

Heckman {\it et al.} (1991) made a spectroscopic study of the extended
gas in a sample of 5 high redshift radio-loud quasars
($z\sim$2-3) (see also Lehnert \& Becker 1998). They found that the kinematical properties are very
similar to the properties of the EELR of HZRG: namely velocity
dispersions across the nebulae consistent with gravitational motions
($<$500 km s$^{-1}$) but with large FWHM$\sim$1000--1500 km
s$^{-1}$. They propose a scenario in which gravitation is at the
origin of these extreme motions: i.e. gas freely falling from
a large distance into the galaxy. An infall process with such
cha\-racteristics could happen during the process of galaxy formation,
if the radio galaxy lies at the bottom of a deep potential well (like a dense cluster).

\item  {\it A group of Ly break galaxies around the radio galaxy.} 

 Recent deep HST images of radio galaxies at $z>$2 show clumpy
and irregular morphologies, consisting of a bright component and a
number of small components, which is suggestive of a merging system
(Pente\-ricci
et al. 1998). Pentericci et al. find that those clumps have similar
characteristics to Ly break gala\-xies and su\-ggest that the host
galaxy of the radio source had itself formed through the merging of such
 smaller
units (see also van Breugel {\it et al.} 1998).

MRC2025-218, MRC2104-242 and SMM02399-0136 have similar morphologies
to the one described above (see optical images in Pentericci {\it et
al.}  1998 and IV98). The presence of several
components is also revealed by our spectra. We also detect  diffuse
emission between (and sometimes beyond) the main clumps which could
have the same nature as the diffuse and asymmetric halos found around
compact clumps in Ly break galaxies (Steidel {\it et al.} 1996).


An important difference between Ly break gala\-xies and the clumps in
HZRG is the large velocities measured {\it in each clump} (FWHM$>$1000
km s$^{-1}$), much larger than the values observed in Ly break
galaxies (FWHM$\leq$200 km s$^{-1}$) ({\it e.g.} Pettini {\it et al.}
1998, Prochaska \& Wolfe 1997).  However, spectroscopy shows that
velocity shifts of $>$1000 km s$^{-1}$ between absorbing and emitting
gas are common in Ly break galaxies. This su\-ggests the presence of
large scale outflows of hundreds km s$^{-1}$ ({\it e.g.} Pettini {\it
et al.} 1998, Franx {\it et al.} 1997), which could be responsible for
large FWHM values if all the gas was ionized. On the other hand no clear 
link has been established between the
absorbing and the ionized gas and, therefore, these velocity
differences might simply occur between unrelated
intervening objects.

\item {\it Bipolar outflows.}

Chambers (1998) has recently proposed that bipolar outflows can be
responsible for the high velocities, morphologies and polarimetric
properties of high redshift radio galaxies. The EELR of HZRG consists
in this model of an expanding bipolar dust shell which scatters light
from a quasar core and has an evacuated interior. 

	Bipolar  outflows can be generated by the
superwind asso\-ciated with a  starburst
in a circumnuclear mole\-cular disk. Evidence for such superwinds has
been found in   Far
Infrared
Galaxies (FIRGs) (Heckman {\it et al.} 1990, H90 hereafter). These galaxies  
show emi\-ssion lines with FWHM of several hundreds km s$^{-1}$ and shifted
by $\sim$1000 km s$^{-1}$ in some objects.
Appropriate superwind models predict outflow velocities of several hundreds km s$^{-1}$. 

	We have calculated some basic parameters characte\-rizing a
superwind which could explain the kinematic properties of the EELR
in HZRG (we want to enphasize the approximate nature of this calculations).   
We have assumed 
a typical radius $r_{neb}$ for
the ioni\-zed nebula
of 20 kpc, a density $n_0$ in the ambient
(undisturbed) medium of  10 cm$^{-3}$ (McCarthy 1993). Two 
emission lines  components of several hundred km s$^{-1}$ 
(unresolved at our spectral
resolution) and shifted by 1000 km s$^{-1}$  will  produce a broad profile
with FWHM$\sim$1000 km s$^{-1}$,  the values observed in our objects. 
Therefore,
we can assume, as concluded for nearby FIRGs, an expanding velocity of 
the nebula $v_{neb}$ of 
several hundreds km s$^{-1}$ (500 km s$^{-1}$). The dynamical time scale 
$t_{dyn}$ for the nebula is given by
equation [7] in H90. We obtain $t_{dyn}\sim$40
Myr, which is in the range  calculated for some  
FIRGs. According to the predictions of the superwind models, this is
comparable to the age of the starburst. For comparison, Dey {\it et al.}
(1997)
derived an upper limit for the young ste\-llar population in the radio galaxy 4C41.17
($z$=3.80) of 600 Myr for a continuous star formation scenario and $\sim$ 16
Myr for an instantaneous starburst model.
  
	On the other hand, if  the total mass of the ionized gas 
(several$\times$10$^8$-10$^9$ M$_{\odot}$, van Ojik 1995, McCarthy
1993)
has been ejected in the outflow and $t_{dyn}\sim$40 Myr, this
implies that the mass injection rate is 7$\leq\frac{dE}{dt}\leq$25 M$_{\odot}$ yr$^{-1}$,
which is also consistent with superwind models for nearby FIRGs.

	We have also calculated the rate of injection of kinetic energy in 
the wind  $\frac{dE}{dt}$, given by eq.
[8] in H90. We obtain
$\sim$3$\times$10$^{47}$ erg s$^{-1}$ which is one order of magnitude higher
than predicted for high power FIRGs.
 This is what we
expect taking into account cosmological evolution of the wind rate
(H90) (the mass injection rate should increase by the same factor, though).
$\frac{dE}{dt}$ is  large compared to  the integrated luminosity  
of the lines (Ly$\alpha$ can
be as luminous as  $\sim$10$^{44}$ erg s$^{-1}$).With a small conversion
efficiency of kinetic energy of the outflow in emission line luminosity,
the outflow could power the nebula. 

	Therefore, a superwind with properties  similar to those
predicted for the high power FIRGs at low redshift, could explain the
kinematic properties of the extended gas in the HZRG.

   The possibility of an outflowing
wind generated in the nuclear region of a powerful radio galaxy is 
 suggested by the nearby radio galaxy Cyg A: a polarization map (Tadhunter
{\it et al.} 1990, Ogle {\it et al.} 1997) shows a biconical structure
suggestive of a dusty reflection nebula which scatters the light of
the hidden active nucleus. The outflow is suggested by the shift
between the na\-rrow emission lines detected in direct light and the
polarized narrow lines (Ogle {\it et al.} 1997). Evidence for a
circumnuclear starburst ring is presented in Fosbury {\it et al.} 1998a.
The velocity shift 
however, is much lower ($\sim$100-200 km s$^{-1}$) than the FWHM values
observed at high $z$.

\end{itemize}

\section{Summary and conclusion}

	We have studied the UV spectra of 3 distant powerful radio galaxies 
and the hyperluminous SMM02399-136 system 
with the goal of understanding the mechanism
responsible for the high velocities observed in the extended gas of
HZRG. 
Large velocities are found in the
extended gas of all the objects (FWHM$>$1000 km s$^{-1}$).

	Interactions between the radio jet and the ambient
gas certainly play a role in some radio galaxies. However, we measure
 high velocities 
in regions where such type of interactions are not taking place and therefore  
other mechanisms must  be at work. 

	Possible explanations for such extreme motions are: 1) 
infall of material
from large distances (gravitational origin). This mechanism could be important
in the process of galaxy formation. 2) A group of Ly break
galaxies in the neighbourhood of the radio galaxy. Large scale outflows in
the individual components are required.  3) Bipolar
outflows produced by superwinds, as observed in nearby FIRGs. 

\begin{acknowledgements}
This work is based on spectroscopic data obtained at La Silla Observatory.
M.Villar-Mart\'\i n acknowledges support of PPARC fellowship to develop
most of this work at the Dept. of Physics and Astronomy in Sheffield (UK). 
Thanks to the referee, Pat McCarthy, for useful comments that contributed
to improve the paper. Thanks also to Raffaella  Morganti for helpful comments on
the radio and optical astrometry of MRC1558-003.
\end{acknowledgements}

\end{document}